\documentclass[10pt]{iopart}
\usepackage{iopams}  
\usepackage[english]{babel}
\expandafter\let\csname equation*\endcsname\relax
\expandafter\let\csname fl\endcsname\relax
\usepackage{amsfonts,amssymb,amstext,amsthm,amscd,amsbsy}
\expandafter\let\csname endequation*\endcsname\relax
\expandafter\let\csname endfl\endcsname\relax
\usepackage{mathtools}
\usepackage[a4paper,top=2.5cm,bottom=2.5cm,left=2.5cm,right=2.5cm]{geometry}
\usepackage{breakcites}
\usepackage{soul}
\usepackage{lineno}
\usepackage{graphicx}
\usepackage{algorithm}
\usepackage{algpseudocode}
\usepackage{multirow}
\usepackage{booktabs}
\usepackage{mathrsfs}
\usepackage{dsfont}
\usepackage{adjustbox}
\usepackage{graphicx}
\usepackage{wrapfig}
\usepackage{indentfirst}
\usepackage{latexsym}
\usepackage{sidecap}
\usepackage{booktabs}
\usepackage{hyperref}

\usepackage{lipsum}
\usepackage{cleveref}
\usepackage[dvipsnames]{xcolor}

\newcommand\myshade{85}
\colorlet{mylinkcolor}{violet}
\colorlet{mycitecolor}{YellowOrange}
\colorlet{myurlcolor}{RoyalBlue}

\hypersetup{
    colorlinks=true,
    linkcolor=myurlcolor,
    citecolor=mycitecolor!\myshade!black,
    filecolor=magenta,      
    urlcolor=magenta,
}

\usepackage{mathrsfs}
\usepackage{textcomp}
\usepackage{braket}
\usepackage{colortbl}
\usepackage{bbm}
\usepackage{comment}
\usepackage{textcomp}
\usepackage{braket}
\usepackage{colortbl}
\usepackage{bbm}

\usepackage{enumitem}
\usepackage{musicography}
\usepackage{caption}
\usepackage{subcaption}

\usepackage{tikz}
\usepackage{pgfplots}
\usetikzlibrary{matrix}
\usepackage{comment}

\usepackage{graphicx}
\usepackage{tikz}
\usetikzlibrary{positioning, fit, arrows.meta, shapes}

\newcommand{\gsm}{MS-GSM$^2\,$}

\newcommand{\marco}[1]{\textcolor{black}{#1}}
\newcommand{\fra}[1]{\textcolor{black}{#1}}
\newcommand{\ema}[1]{\textcolor{black}{#1}}

\usepackage{fancyhdr}
\usepackage{lipsum}
\usepackage{comment}
\numberwithin{table}{section}
\usepackage{multirow}
\newcolumntype{?}{!{\vrule width 3pt}}

\begin{document}


\title{A multiscale radiation biophysical stochastic model describing the cell survival response at ultra-high dose rate}

\author{Marco Battestini$^{1,2}$, Marta Missiaggia$^{2,3}$, Sara Bolzoni$^{1,2}$, Francesco G. Cordoni$^{2,4,a,*}$ and Emanuele Scifoni$^{2,a}$}

\renewcommand{\thefootnote}{\fnsymbol{footnote}}
\footnotetext{{\scriptsize $^{1}$ Department of Physics, via Sommarive 14, 38123, Trento, Italy}}
\footnotetext{{\scriptsize $^{2}$ Trento Institute for Fundamental Physics and Application (TIFPA), via Sommarive 15, 38123, Trento, Italy}}
\footnotetext{{\scriptsize $^{3}$ Department of Radiation Oncology, University of Miami Miller School of Medicine, 33136, Miami FL, USA}}
\footnotetext{{\scriptsize $^{4}$ Department of Civil, Environmental and Mechanical Engineering, via Mesiano 77, 38123, Trento, Italy}}
\footnotetext{{\scriptsize $^{a}$ equal contribution}}
\footnotetext{{\scriptsize $^{*}$ corresponding author}}

\footnotetext{{\scriptsize E-mail addresses: marco.battestini@unitn.it, marta.missiaggia@miami.edu, sara.bolzoni@studenti.unitn.it, francesco.cordoni@unitn.it, emanuele.scifoni@tifpa.infn.it}}
\vspace{10pt}

\begin{abstract}
Ultra-high dose-rate (UHDR) radiotherapy, characterized by an extremely high radiation delivery rate, represents one of the most recent and promising frontier in radiotherapy. UHDR radiotherapy, addressed in the field as FLASH radiotherapy, is a disruptive treatment modality with several benefits, including significantly shorter treatment times, unchanged effectiveness in treating tumors, and clear reductions in side effects on normal tissues. While the benefits of UHDR irradiation have been well highlighted experimentally, the biological mechanism underlying the FLASH effect is still unclear and highly debated. Nonetheless, to effectively use UHDR radiotherapy in clinics, understanding the driving biological mechanism is paramount. Since the concurrent involvement of multiple scales of radiation damage has been suggested, we developed the MultiScale Generalized Stochastic Microdosimetric Model (MS-GSM$^2$), a multi-stage extension of the GSM2, which is a probabilistic model describing the time evolution of the DNA damage in an irradiated cell nucleus. The MS-GSM$^2$ can investigate several chemical species combined effects, DNA damage formation, and time evolution. We demonstrate that the MS-GSM$^2$ can predict various \textit{in-vitro} UHDR experimental results across various oxygenation levels, radiation types, and energies. The MS-GSM$^2$ can accurately describe the empirical trend of dose and dose rate-dependent cell sensitivity over a wide range, consistently describing multiple aspects of the FLASH effect and reproducing the main evidence from the \textit{in-vitro} experimental data. Our model also proposes a consistent explanation for the differential outcomes observed in normal tissues and tumors, \textit{in-vivo} and \textit{in-vitro}.
\end{abstract}

\section{Introduction}
A current frontier in the treatment of cancer is \ema{undoubtely} the so-called FLASH \ema{radio}therapy, a revolutionary radiation therapy technique based on the ultra-high dose rate (UHDR) irradiation, i.e., a faster beam delivery (total irradiation time $<$ 100 ms and a mean dose rate of the total irradiation $>$ 40 Gy/s for a single high dose, usually higher than 10 Gy) compared to conventional (Conv) irradiation (dose rate of 0.03 - 0.1 Gy/s) \cite{Vozenin2022, LimoliVozenin2023}. The peculiarity of this novel approach is that it allows to significantly reduce the side effects to healthy tissues \cite{Vozenin2019, Montay-Gruel2017, Montay-Gruel2019, Loo2017}, with the same tumor efficacy \cite{Favaudon2014}. This phenomenon is known as the FLASH effect, and it has been showed on \textit{in-vitro} cells \cite{Adrian2020, Tessonnier2021, Tinganelliinvitro2022}, on animals \cite{Vozenin2019, Montay-Gruel2017, Montay-Gruel2019, Loo2017, Montay-Gruel2021, Diffenderfer2020, Cunningham2021, Tinganelliinvivo2022, Sorensen2022} and also on humans \cite{Bourhis2019}. From a clinical point of view \cite{Daugherty2023}, \fra{FLASH radiotherapy allows} \marco{to improve the quality of radiotherapy treatment significantly.}  

Despite clear experimental evidence supporting the FLASH effect, highlighting the benefits of the UHDR irradiations, the biological mechanism underlying the FLASH effect is still unclear and highly debated \cite{LimoliVozenin2023}. 
In recent years, several \ema{tentative} hypotheses \cite{Spitz2019, Labarbe2020, Wardman2020, Zhou2020, Pratx2020, Ramos2020, Zakaria2020, abolfath2020oxygen, petersson2020quantitative} have been proposed to explain the FLASH effect \cite{Friedl2022}, many of which have been \ema{considered implausible following} further experimental evidence, e.g. \cite{Pratx2020, Ramos2020}. None have yet been able to explain it fully.\\
However, what has emerged from all \ema{these efforts} is that multiple scales of radiation damage \ema{action should be|} involved \cite{Weber2022}. The crucial role of \fra{chemical environment and of the redox system} has been underlined \cite{Labarbe2020, Weber2022, Wardman2020, Favaudon2022}. 

\fra{Over the last years, a plethora} of mechanistic models have been proposed and developed \cite{Labarbe2020, abolfath2020oxygen, liew2021deciphering, petersson2020quantitative, abolfath2022differential, Zhu2021, Spitz2019, Hu2021, Alanazi2020, Zakaria2020, Pratx2020}. Most such models \ema{have been} based on \ema{the hypothesis of a single driving mechanism} \marco{and they} can only partially reproduce the experimental data, suggesting \marco{some limits in their capability to explain and predict the mechanism of the effect of the FLASH effect}. \ema{Instead,}\marco{We hypothesize that the FLASH biological effect is \marco{due to} a interplay of various spatio-temporal scales, as suggested by \cite{Weber2022}.}

\begin{figure*}
\centering
\includegraphics[width=.75\linewidth]{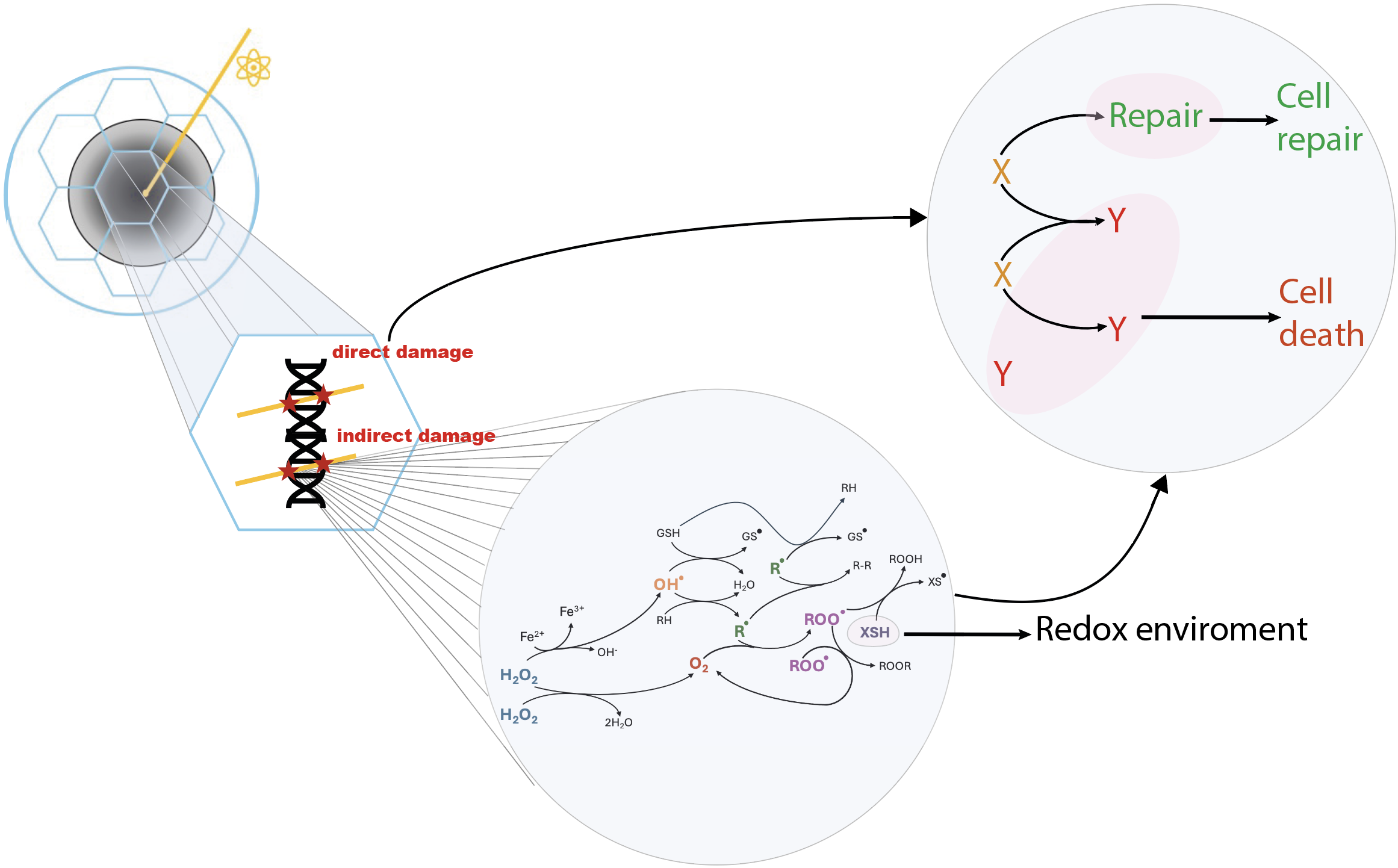}
\caption{Schematic representation of the MS-GSM$^2$ workflow.}
\label{fig:scheme}
\end{figure*}

\marco{In this context,} we develop the \textit{MultiScale Generalized Stochastic Microdosimetric Model} (MS-GSM$^2$), a multi-stage extension of the \textit{Generalized Stochastic Microdosimetric Model} (GSM$^2$) \cite{cordoni2021generalized, cordoni2022cell, cordoni2022multiple, cordoni2023spatial, missiaggia2022cell, Battestini2023}. The GSM$^2$ is a stochastic Markov model that describes the formation and repair of radiation-induced DNA damages in a cell nucleus. The GSM$^2$ can reproduce \textit{in-vitro} cell survival data for conventional irradiation for different particles \cite{missiaggia2022cell, cartechini2024integrating}. This work aims to establish a general mathematical framework seamlessly integrating various stochastic spatial and temporal scales to study the combined effects of multiple chemical species and the formation and progression of DNA damage within the UHDR regime. These scales range from the nearly instantaneous ionization of biological tissue to the rapid chemical environment contributing to DNA damage formation and, ultimately, to the slow biological repair processes determining a cell's fate. \fra{The hypothesis that UHDR modifies the chemical environment to induce a reduction of indirect damage yield, with a consequent tissue sparing. The reduction of DNA damage is supported by experimental evidence, \cite{Tessonnier2021, Cooper2022, Wanstall2024}.} In this work, we fulfill the development and validation of \gsm \cite{Battestini2023}. We first include the reduction of radiation-induced DNA damage owing to lower oxygen concentration, a well-documented effect in radiobiology\ema{ with a strong interplay with the radiation quality of the beam} \cite{Scifoni2013}. Unlike existing models, the \gsm explicitly incorporates oxygen concentration through a term in the system of differential equations that govern its time-dependent evolution influenced by environmental reactions. \fra{Further, we significantly restructure and enhance the chemical reaction network previously considered in \cite{Labarbe2020}. The water radiolysis reactions, which occur \ema{during the heterogeneous chemical stage, thus} at faster time scales than those of organic chemical species, which are the main species of interest in the present framework, are now included through a parametrization process using the Monte Carlo toolkit TRAX-CHEM \cite{Boscolo2018}.
Consequently, we derive a chemical network of ten reactions, represented by five ordinary differential equations (ODEs) that describe the chemical environment. The influence of radiation is integrated into the chemical network, characterized by stochastic random arrival times. Ultimately, the chemical environment is modeled using piecewise deterministic ODEs with random jumps in amplitude at random intervals. This captures the stochastic nature of radiation energy deposition in time and space. This chemical reaction network is then coupled to the Markov model as outlined by GSM$^2$, incorporating DNA damage repair mechanisms, culminating in a comprehensive multi-scale stochastic mathematical model.}

\marco{We develop a versatile mathematical model capable of describing the intricate mechanisms underlying the FLASH effect while maintaining flexibility for predictive modeling, being, ultimately, its purpose \ema{dedicated to} replicate experimental outcomes.} \fra{We implement such a model by an efficient coupling of \textit{Stochastic Simulation Algorithm} (SSA) to describe the radiation random arrival times and the Markov model with the solution of the ODE. This approach provides accurate predictions and ensures swift implementation of a complex system.} We successfully replicate the main \textit{in-vitro} cell-survival UHDR experimental data from the main studies available in literature \cite{Adrian2020, Tessonnier2021, Tinganelliinvitro2022}, encompassing a diverse spectrum of particle types, energies,\ema{delivered doses} and physiological conditions. The accuracy of our predictions across this extensive range of physical, chemical, and biological conditions provides compelling evidence supporting \ema{that the \gsm grasps all the physics behind the } potential mechanism underlying the FLASH effect. \fra{Supported by the \textit{in-silico} prediction of the \gsm, we eventually propose that the different redox environment between tumor and healthy tissue is at the origin of the differential effect seen on the two different tissues.}

\section*{Results}

\subsection*{The model}\label{SEC:Model}

The GSM$^2$ \cite{cordoni2021generalized, cordoni2022cell} is a Markov model that generalizes the standard Poisson statistics for the number of radiation-induced DNA damages \cite{cordoni2022multiple, cordoni2023emergence}. Specifically, the GSM$^2$ encompasses the randomness of radiation energy deposition, DNA damage formation, and the temporal repair of these damages within a singular framework. The GSM$^2$ is considerably extended in the present work to include the effect of UHDR into the model via a suitable chemical reaction network, developing a multi-stage tool, the MS-GSM$^2$. \fra{In the following paragraphs, we outline the mathematical framework of MS-GSM$^2$, dividing the model's components into stages, from the rapid physical stage to the slow biological stage.} \marco{The basic idea of the model is also schematized in Figure \ref{fig:scheme}.}

\subsubsection*{The physical stage}\label{SEC:PhysS}

We describe a cell nucleus using a cylinder of radius $R_N = 8\,\mu$m ($xy$-plane) and length $2 R_N$ ($z$-direction), divided into $N_d = 52$ cubic domains. \fra{The absorbed dose of every energy deposition event $z$ is distributed on the $N_d$ domains of the $ xy$ plane of the cell nucleus, according to the Amorphous Track (AT) model described in \cite{Scholz1996}. The AT model hypothesizes that the energy deposition by a particle has \ema{cylindrical} symmetry, characterized by a uniform energy distribution over a circle with radius $R_c$ centered on the impact point, and decreases proportionally to $r^{-2}$, where $r$ is the distance from the impact point. Beyond a certain distance $R_p$, no energy is deposited, i.e.}
\begin{equation}\label{EQN:DcDp}
D(r) =
\begin{cases}
\frac{1}{\pi R_c^2} \left( \frac{\mathrm{LET}}{\rho \left( 1 + 2 \ln \left( R_p/R_c \right) \right)} \right) & r \leq R_c\,,\\
\frac{1}{\pi r^2} \left( \frac{\mathrm{LET}}{\rho \left( 1 + 2 \ln \left( R_p/R_c \right) \right)} \right) & r \in [R_c,R_p]\,,\\
\end{cases}
\end{equation}
\marco{where $\mathrm{LET}$ is the linear energy transfer, which characterizes the radiation quality, and $\rho$ is the water density.} In this study, $R_p$ is defined as the radius at which $80\%$ of the dose is deposited by the track, referred to as $R_{80}$ \cite{Weber2022, Battestini2023}. \ema{This approximation neglects the residual extension of the dose until the maximum distance allowed to the secondary electrons propagation, which is spread along a broader circular crown}. The impact position of the primary particles is sampled randomly from a uniform distribution on a circle of radius $R_N + R_{80}$ on the $xy$-plane. \fra{The energy deposited in each domain is then calculated, demonstrating the stochastic nature of energy deposition, as different impact points result in varying amounts of energy being deposited in the cell nucleus and, consequently, in each domain.}

\fra{All of the energy} deposition events up to the end of the irradiation, i.e., $T_{irr} = D / \dot{D}$, where $D$ is the imparted macroscopic dose, and $\dot{D}$ is the mean dose rate, \fra{are calculated obtaining} the total number of simulated particles. The elapsed time $\tau$ between two consecutive deposition events is sampled from an exponential distribution, which depends on the dose rate and the mean energy deposited in a single event.

Lastly, in this work, we generalize the dose-delivery simulation to describe any time structure. In particular, in the case of ions, we consider a unique pulse described by \marco{a macroscopic dose prescription $D$ and an overall mean dose rate $\dot{D}$, and we simulate the discrete energy deposition events of this pulse, according to the previous description. On the contrary, for electrons, according to the experimental dose delivery time structure, we simulate irradiation characterized by a sequence of an integer number of pulses, identified by a dose-per-pulse and a pulse time, interspersed with pauses, defined by a pulse repetition frequency. As before, we sample the energy deposition events for each pulse.}

\subsubsection*{The chemical stage}

\fra{We consider a chemical reaction network of nine reactions described by five ODEs. The derived system is a severe optimization of the one proposed in} \cite{Labarbe2020}, on the one hand by simplifying some equations, cutting away some negligible reactions, and on the other hand by adjusting some reaction constants, among all of those listed in Sections S.1 and S.2 of \cite{Labarbe2020}.

In particular, in \cite{Labarbe2020}, the reactions involving only chemical species derived from the water radiolysis, such as $\mathrm{e^{-}_{aq}}$, $\mathrm{O\mathrm{H^{\bullet}}}$, $\mathrm{H^{\bullet}}$, $\mathrm{H_2}$, $\mathrm{O_2^{\bullet -}}$, have much lower rates than both the reactions between these species with organic \ema{molecules}, and those between organic species alone. Indeed, it is possible to neglect these types of reactions, thus separating the two timescales: a faster one which we assume to be instantaneous, i.e., the \ema{early heterogeneous stage of water radiolysis} that is incorporated through the production terms \fra{as described by G-values} of chemical species in the system of equations\marco{, i.e., the number of chemical species produced per 100 eV of energy absorbed by the medium}, and that of the reactions involving the various organic chemical species, such as $\mathrm{RH}$, $\mathrm{R^{\bullet}}$, $\mathrm{ROO^{\bullet}}$, vitamin E, thiols and their derivatives, described explicitly by the reaction rates. \fra{It is worth stressing that \ema{the latter} species is the one \ema{which we directly link} to the radiation-induced damage formation.} In this way, \fra{we obtain, for each domain $d = 1,\dots, 52$, the following \ema{system}}
\begin{equation}\label{EQN:chemODEs}
\begin{cases}
        \frac{d}{dt} [\mathrm{O_2}]^d (t) &=
        k_{1}\cdot[\mathrm{R^\bullet}]^d (t)\cdot[\mathrm{O_2}]^d (t) + k_{2}\cdot[\mathrm{ROO^\bullet}]^d (t) \cdot[\mathrm{ROO^\bullet}]^d (t) + k_{3}\cdot \frac{[\mathrm{cat}]}{k_m+[\mathrm{H_2} \mathrm{O_2}]^d (t)}\cdot[\mathrm{H_2} \mathrm{O_2}]^d (t)\,, \\

        \frac{d}{dt} C_{\mathrm{H_2} \mathrm{O_2}}^d (t) &=
        k_{4}\cdot[Fe^{2+}]\cdot[\mathrm{H_2} \mathrm{O_2}]^d (t) - 2k_{3}\cdot \frac{[\mathrm{cat}]}{k_m+[\mathrm{H_2} \mathrm{O_2}]^d (t)}\cdot[\mathrm{H_2} \mathrm{O_2}]^d (t) + \mathrm{G}_{\mathrm{H_2} \mathrm{O_2}} \rho \dot{z}^d\,, \\

        \frac{d}{dt} C_{\mathrm{OH^\bullet}}^d (t) &= - k_{5}\cdot[\mathrm{RH}]\cdot[\mathrm{OH^\bullet}]^d (t) + 
        k_{4}\cdot[Fe^{2+}]\cdot[\mathrm{H_2} \mathrm{O_2}]^d (t) - k_{6}\cdot[\mathrm{XSH}]\cdot[\mathrm{OH^\bullet}]^d (t)
        + \mathrm{G}_{\mathrm{OH^\bullet}} \rho \dot{z}^d\,, \\

        \frac{d}{dt} [\mathrm{R^\bullet}]^d (t) &=
        k_{5}\cdot[\mathrm{RH}]\cdot[\mathrm{OH^\bullet}]^d (t) - k_{1}\cdot[\mathrm{R^\bullet}]^d (t)\cdot[\mathrm{O_2}]^d (t) - k_{7}\cdot[\mathrm{XSH}]\cdot[\mathrm{R^\bullet}]^d (t) +\\ & - 2k_{8}\cdot[\mathrm{R^\bullet}]^d (t) \cdot[\mathrm{R^\bullet}]^d (t)
        + \mathrm{G}_{\mathrm{R^{\bullet}}} \rho \dot{z}^d\,, \\

        \frac{d}{dt} [\mathrm{ROO^\bullet}]^d (t) &=  
        k_{1}\cdot[\mathrm{R^\bullet}]^d (t)\cdot[\mathrm{O_2}]^d (t) -k_{9}\cdot[\mathrm{XSH}]\cdot[\mathrm{ROO^\bullet}]^d (t) - 2k_{2}\cdot[\mathrm{ROO^\bullet}]^d (t) \cdot[\mathrm{ROO^\bullet}]^d (t)\,,        
\end{cases}
\end{equation}
where $[\mathrm{H_2} \mathrm{O_2}]^d (t) = \frac{C_{\mathrm{H_2} \mathrm{O_2}}^d (t)}{1+\frac{K_{H_2O_2}}{[H^+]}}$ and $[\mathrm{OH^\bullet}]^d (t) = \frac{C_{\mathrm{OH^\bullet}}^d (t)}{1+\frac{K_{OH^\bullet}}{[H^+]}}$, that describe the acid-base equilibria, in analogy to \cite{Labarbe2020}, $\rho$ is the water density, \marco{while $\mathrm{G}_{\xi}$ are the G-values, according to TRAX-CHEM simulations \cite{Boscolo2018, Boscolo2020}} \ema{, computed at 1 $\mu s$.} Overall, we have a system of 260 ODEs.

As previously mentioned, we update the values of some reaction constants within the ranges we found in the literature to describe the experiments better, as shown in the next section. In particular, in this work, we consider $k_{1} = 10^8 (mol/l)^{-1} \cdot s^{-1}$ (typical range of $[10^6, 10^{10}] (mol/l)^{-1} \cdot s^{-1}$ \cite{Labarbe2020, Spitz2019, Neta1990, Epp1976, Michael1986}), $k_{2} = 10^5 (mol/l)^{-1} \cdot s^{-1}$ (typical range of $[10^3, 10^5] (mol/l)^{-1} \cdot s^{-1}$ \cite{Labarbe2020, Riahi2010}), and $k_{9} = 10^2 (mol/l)^{-1} \cdot s^{-1}$ \cite{Spitz2019, Wardman2020}. With this last modification, we include the scavenging role of $\mathrm{GSH}$ to take into account the elimination of $\mathrm{ROO^\bullet}$ from the environment, which had been partially neglected in \cite{Labarbe2020}, according to \cite{Wardman2020, Tan2023}. \fra{Additionally, the terms $\dot{z}$ represent the \ema{local} energy deposition within the domain, as outlined in the preceding section. Specifically, the terms $\dot{z}$ are characterized by random impulses in both time and magnitude, which depict the stochastic nature of the time delivery structure.}

Table S1 of the SI Appendix reports the reaction rate constants' values and the chemical species' initial concentrations considered in this work.

\subsubsection*{The bio-chemical stage}

In the biochemical stage, the deposition of radiation energy and the chemical environment are interconnected with the yield of biological damage.

The peculiarity of the GSM$^2$ is that it overcomes the usual assumptions of a Poissonian distribution of the number of radiation-induced DNA damage, which limits the possible predictions of biological endpoints\ema{, as demonstrated in} \cite{missiaggia2022cell}. The GSM$^2$ assumes the possibility of generation of two types of DNA damage, the lethal lesion $Y$, i.e., irreparable damage, leading to cell inactivation, and sub-lethal lesion $X$, i.e., repairable damage, where $Y$ and $X$ are two $\mathbb{N}-$valued random variables.

A protracted dose-rate term $\dot{d}$ modulates the energy deposition events and subsequent lesion creation \cite{cordoni2021generalized}. The next treatment made several generalizations regarding the creation of DNA damage. The average damage yield consequent to an energy deposition $z$ is calculated as
\begin{equation}\label{EQN:kDNA}
\kappa\left( \mathrm{LET}, [\mathrm{O_2}] \right) \,z = \frac{\text{DSB($\mathrm{LET}$)}}{\mathrm{OER} \left( \mathrm{LET}, [\mathrm{O_2}] \right) } \,z\,.
\end{equation}
Above, $\text{DSB($\mathrm{LET}$)}$ is the analytical formula representing the average \ema{yield of damage} obtained via track‑structure simulations in \cite{kundrat2020analytical}, whereas $\mathrm{OER} \left( \mathrm{LET}, [\mathrm{O_2}] \right)$ represents the Oxygen Enhancement Ratio (OER), i.e., the ratio of doses giving the same biological effect in the presence of a given concentration of oxygen $[\mathrm{O_2}]$ as compared to the full oxygenation (normoxia), characterizes a well-documented experimental phenomenon in radiation biophysics where damage fixation increases at higher oxygen concentrations, leading to enhanced radiation lethality. The OER is described in \cite{Scifoni2013, Epp1972} as
\begin{equation}\label{EQN:OER}
\mathrm{OER} \left( \mathrm{LET}, [\mathrm{O_2}] \right) = \frac{K_{\mathrm{O_2}} \cdot  \frac{K_{\mathrm{LET}} \cdot M + \mathrm{LET}^\gamma}{K_{\mathrm{LET}} + \mathrm{LET}^\gamma} + [\mathrm{O_2}]}{K_{\mathrm{O_2}} + [\mathrm{O_2}]} \,,
\end{equation}
where $M$ represents the maximum effect, $K_{\mathrm{O_2}}$ is the flex point for $[\mathrm{O_2}]$, as outlined in \cite{Epp1972}, $K_{\mathrm{LET}}$ denotes the flex point for $\mathrm{LET}$, that is, the average energy deposited per unit path length, which characterizes the radiation quality of the ionizing particle, and $\gamma$ is an additional fitted parameter, as emphasized in \cite{Scifoni2013}. To consider the impact of the chemical environment, the damage yield is divided into a \textit{direct contribution}, which describes the \ema{lesions formed} immediately from energy deposition regardless of the chemical environment, and an \textit{indirect contribution} resulting from the action of radicals and other molecular species. We assume that following an energy deposition event $z$, on average, a proportion $(1-q) \kappa\left( \mathrm{LET}, [\mathrm{O_2}] \right)$ yield a direct damage. In contrast, the remaining $q \kappa\left( \mathrm{LET}, [\mathrm{O_2}] \right)$ will yield and indirect damage. These proportions are again calculated using the formulas given in \cite{kundrat2020analytical}. Being the direct damage independent of the chemical environment, we describe the number of induced lesions with a Poisson random variable of average $(1-q) \kappa\left( \mathrm{LET}, [\mathrm{O_2}] \right)\,z$. Regarding indirect damages, we assume that the number of such damages depends on the chemical environment and the irradiation pattern. We assume that the concentration of peroxyl radical $[\mathrm{ROO^\bullet}]$ is directly linked to the number of indirect damage created, so we describe this process with a Poisson random variable of average $q \kappa\left( \mathrm{LET}, [\mathrm{O_2}] \right)\varrho \int_{0}^t [\bar{\mathrm{ROO^\bullet}}](s) \,ds$. \marco{This quantity is normalized in such a way to obtain the same average of damage yield computed with track‑structure simulations in \cite{kundrat2020analytical} at standard conditions, i.e., conventional irradiation and 21 \% $\mathrm{O_2}$. At the same time, it decreases as the oxygenation level decreases and the dose rate increases. Additional information about this can be found in the SI Appendix.}

Overall, the damage yield creation can be summarized by the following network
\begin{equation}\label{EQN:Reactd}
\begin{split}
&  X \xrightarrow{\dot{d}}
\begin{cases}
X + Z_{X;\mathrm{dir}} & \mbox{with probability} \quad 1-q \in [0,1]\,,\\
X + Z_{X;\mathrm{indir}} & \mbox{with probability} \quad q \in [0,1]\,,\\
\end{cases}\,.\\
\end{split}
\end{equation}
where $ Z_{X;\mathrm{dir}}$ and $Z_{X;\mathrm{indir}}$ are two Poisson random variables, with average, respectively, $\kappa\left( \mathrm{LET}, [\mathrm{O_2}] \right)z$ and $\kappa\left( \mathrm{LET}, [\mathrm{O_2}] \right)\varrho \int_{0}^t [\bar{\mathrm{ROO^\bullet}}](s) \,ds$. These two processes will be described via two unitary Poisson jump processes $\mathrm{P}^{(c,d)}_{\mathrm{dir}}$ and $\mathrm{P}^{(c,d)}_{\mathrm{indir}}$, respectively. A completely analogous argument has been used to form lethal lesions $Y$, with the function $\kappa$ substituting by $\lambda = 10^{-3} \kappa$.

The formulation described above offers a comprehensive overview of the process of radiation-induced damage formation, distinguishing between direct and indirect lesions. It introduces \ema{for the first time} an explicit dependency on the chemical environment through the OER formulation \eqref{EQN:OER} and explicitly incorporates radicals' impact on damage yield. Consequently, as the pattern of radiation energy deposition significantly influences this factor, the formulation enables us to depict the dual dependency of damage yield on oxygenation and the presence of organic radicals, demonstrating a sparing effect at progressively higher dose rates, \marco{as \ema{shown} experimentally by \cite{Cooper2022, Wanstall2024}.}

\subsubsection*{The biological stage}

It is possible to describe the state of the system at time $t$ considering two random variables $\left (Y(t),X(t)\right )$, i.e., the number of lethal and sub-lethal lesions at time $t$, respectively. In particular, a sub-lethal lesion can undergo three different pathways,
\begin{equation}\label{EQN:React}
X \xrightarrow{a} Y\,,\quad X \xrightarrow{r} \emptyset\,,\quad X + X \xrightarrow{b} Y\,,\\
\end{equation}
where $a$ is the rate of the evolution of a sub-lethal lesion to a lethal one, $b$ describes the rate of pairwise interaction of two sub-lethal lesions, while $r$ represents the repair rate, with $\emptyset$ the ensemble of healthy cells. More details about the mathematical framework of this model can be found in \cite{cordoni2021generalized, cordoni2022cell}. These processes are described by $\mathrm{P}^{(c,d)}_h$, with $h \in \{a,b,r\}$, that are independent unitary Poisson jump-processes acting on a single domain, \cite{Wei}.

In the end, we obtained the following system of equations for the MS-GSM$^2$ for each domain $d =1,\dots, N_d$,
\begin{equation}\label{EQN:MultiGSM}
\begin{cases}
Y^{(c,d)}(t) &= \mathrm{P}^{(c,d)}_a\left ( \int_0^t a X^{(c,d)}(s)ds\right )+\mathrm{P}^{(c,d)}_b\left (\int_0^t b X^{(c,d)}(s)(X^{(c,d)}(s)-1)ds\right ) +\\&+ Z_{Y;\mathrm{dir}}^{(c,d)}\mathrm{P}^{(c)}_{Y; \mathrm{dir}}\left ((\dot{d} \,t\right ) + Z_{Y;\mathrm{indir}}^{(c,d)}\mathrm{P}^{(c,d)}_{Y; \mathrm{indir}}\left (\varrho \int_{0}^t [\bar{\mathrm{ROO^\bullet}}](s) \,ds \right )\,,\\
X^{(c,d)}(t) &=  - \mathrm{P}^{(c,d)}_a\left (\int_0^ta X^{(c,d)}(s)ds\right ) - \mathrm{P}^{(c,d)}_r\left (\int_0^t r X^{(c,d)}(s)ds\right ) +\\
&- 2\mathrm{P}^{(c,d)}_b\left (\int_0^t b X^{(c,d)}(s)(X^{(c,d)}(s)-1) ds\right ) +Z_{X;\mathrm{dir}}^{(c,d)}\mathrm{P}^{(c)}_{X; \mathrm{dir}}\left (\dot{d} \, t\right )+\\ &+ Z_{X;\mathrm{indir}}^{(c,d)}\mathrm{P}^{(c,d)}_{X; \mathrm{indir}}\left (\varrho \int_{0}^t [\bar{\mathrm{ROO^\bullet}}](s) \,ds \right )\,,\\
\frac{d}{dt}\xi^d(t) &= f_{\xi^d} \left (\xi^d(t)\right ) + \mathrm{G}_{\xi^d} \rho \dot{z}^d\,,\\ 
\end{cases}
\end{equation}
where the last equation \ema{represents} the random ODE system \eqref{EQN:chemODEs} describing the chemical environment. The first two equations in \eqref{EQN:MultiGSM} is a pathwise description in terms of jump-type \textit{Stochastic Differential Equations} (SDE) of the processes (\ref{EQN:React}) and (\ref{EQN:Reactd}); a similar pathwise formulation of the GSM$^2$, with no explicit chemical environment formulation, has also been derived in \cite{cordoni2023spatial}, we refer to \cite[Chapter 13]{Wei}.

\begin{figure*}[!t]
     \centering
     \begin{subfigure}[b]{0.6\textwidth}
         \centering
         \includegraphics[width=\linewidth]{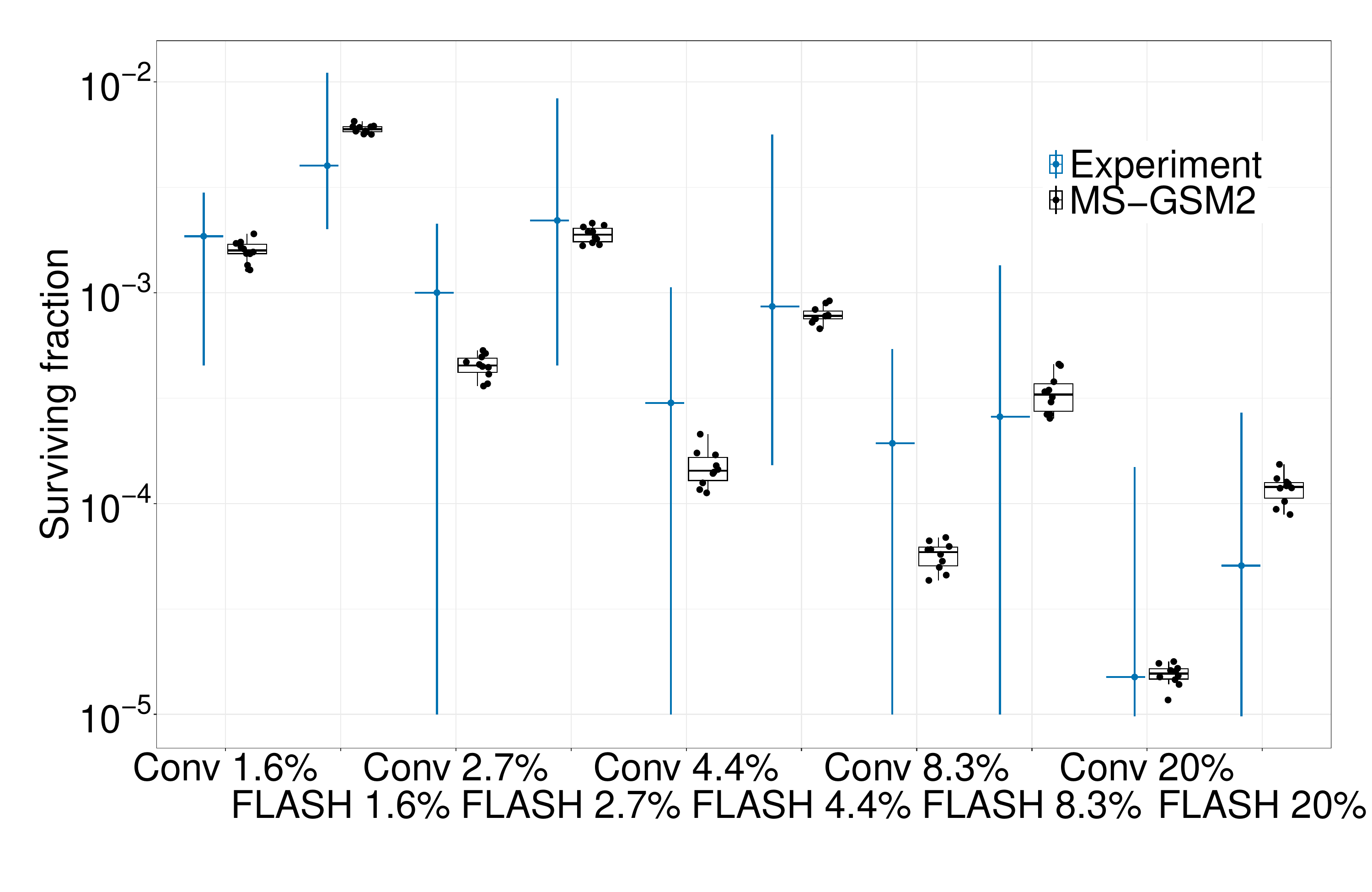}
         \caption{DU145 cell survival in the case of 10 MeV electrons irradiation, at 18 Gy for both Conv (14 Gy/min) and FLASH (600 Gy/s) dose rates, for different oxygenations (1.6\%, 2.7\%, 4.4\%, 8.3\%, 20\% $\mathrm{O_2}$).}
         \label{fig:electron}
     \end{subfigure}
     \hfill
          \begin{subfigure}[b]{0.45\textwidth}
         \centering
         \includegraphics[width=\linewidth]{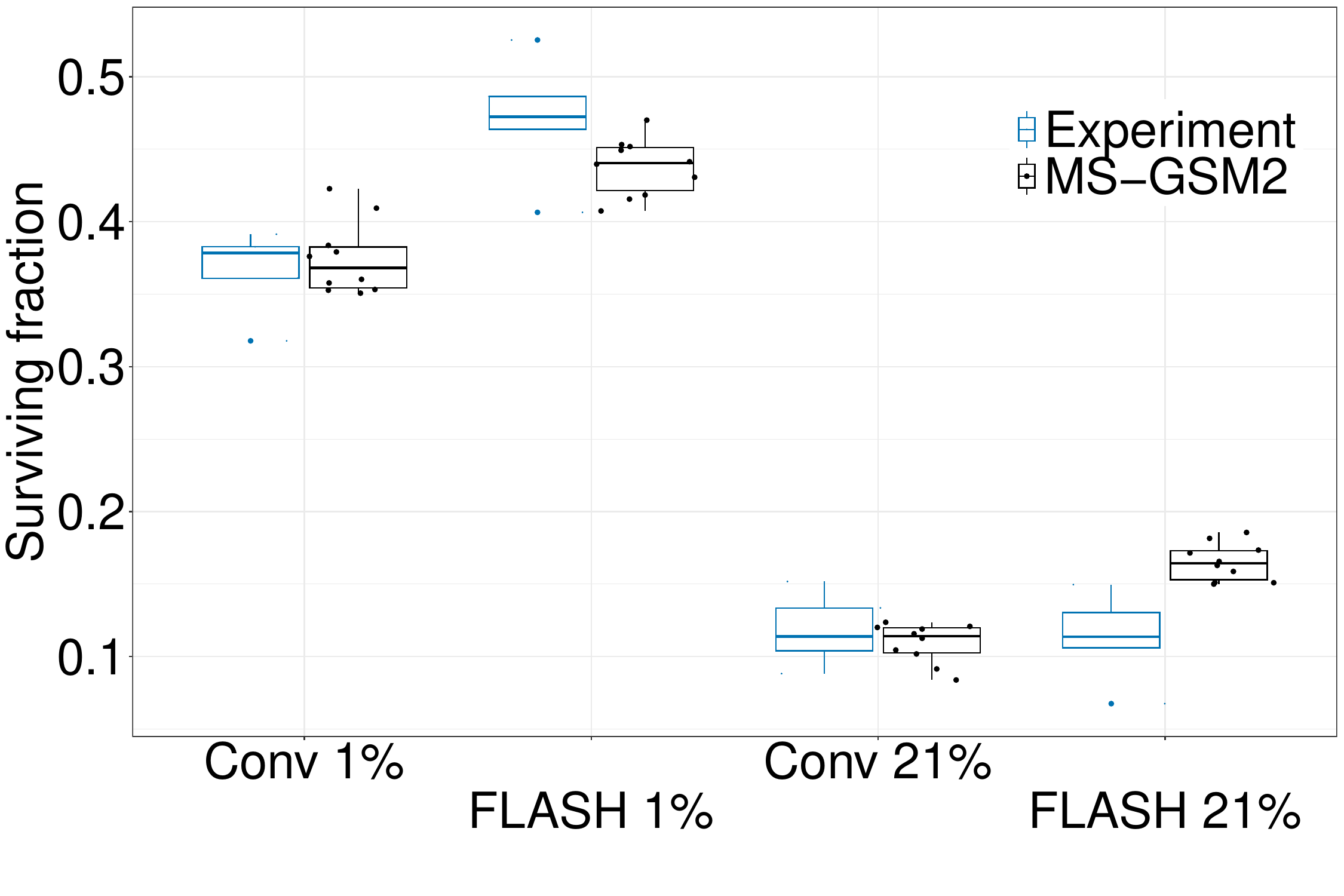}
         \caption{A549 cell survival in the case of 4.5 keV/$\mu$m helium ions irradiation, at 8 Gy for both Conv (0.12 Gy/s) and FLASH (205.13 Gy/s) dose rates, for different oxygenations (1\%, 21\% $\mathrm{O_2}$).}
         \label{fig:heliumd8}
     \end{subfigure}
     \begin{subfigure}[b]{0.45\textwidth}
         \centering
         \includegraphics[width=\linewidth]{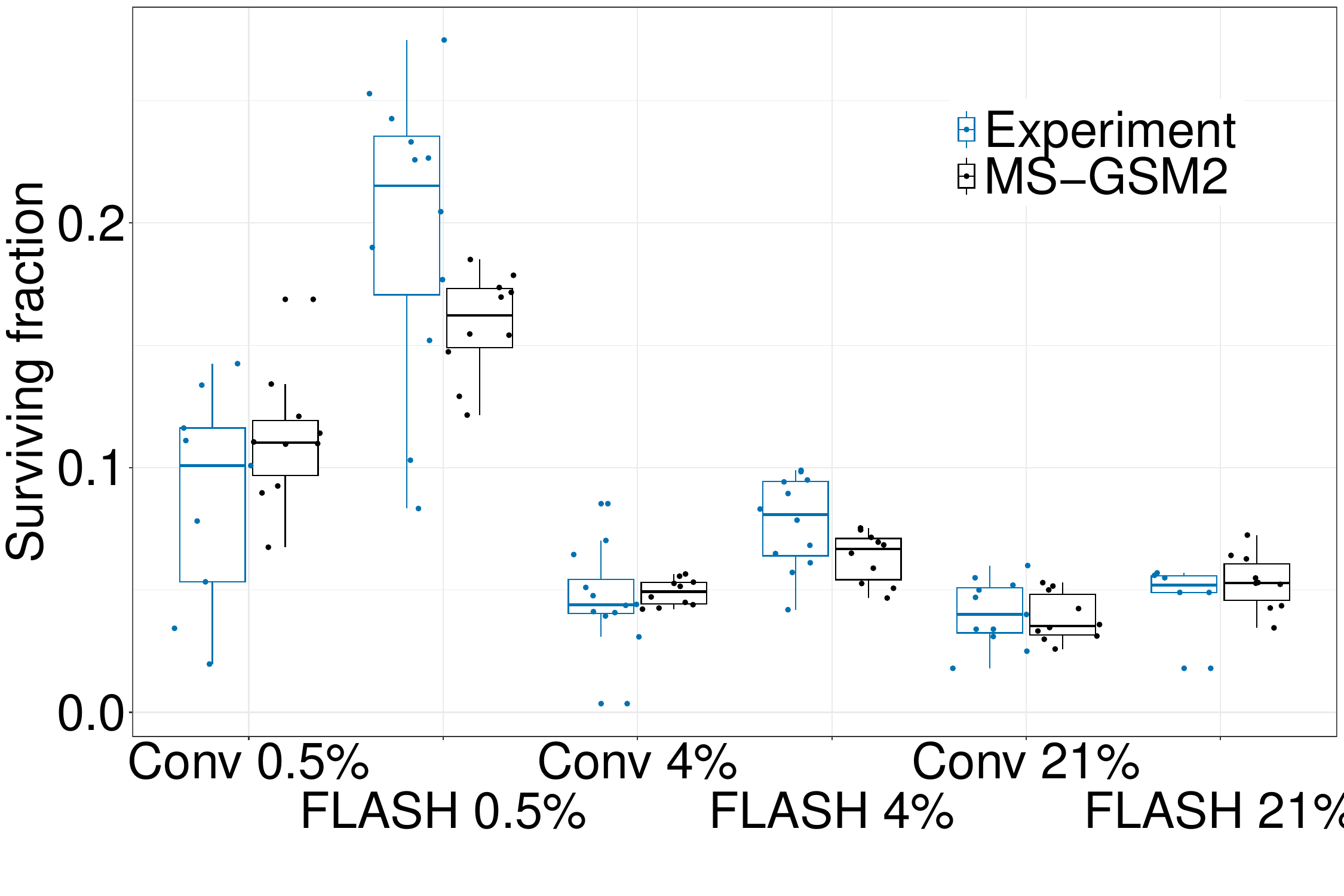}
         \caption{CHO-K1 cell survival in the case of 13 keV/$\mu$m carbon ions irradiation, at 7.5 Gy for both Conv (0.6 Gy/s) and FLASH (70 Gy/s) dose rates, for different oxygenations (0.5\%, 4\%, 21\% $\mathrm{O_2}$).}
         \label{fig:carbon}
     \end{subfigure}
     \hfill
     \caption{Comparison of the MS-GSM$^2$ results with experimental data.}
     \label{fig:electron_carbon}
\end{figure*}

We emphasize that the enhancements in this work render the MS-GSM$^2$ highly flexible and comprehensive, enabling the examination of various scenarios: physically, it accommodates any radiation dose-delivery \ema{complex} time structure, type, and energy without simplification on the dose rate structures as done in all the existing mechanistic \ema{models} where an average macroscopic dose rate is \ema{mostly} assumed; chemically, it considers all chemical and environmental conditions, such as oxygenation and scavenger concentrations; biologically, it describes multiple effects, from well-established ones like the oxygen effect on DNA damage fixation during conventional irradiation, to widely debated ones like the sparing effect seen at UHDR regimes across various cell lines. The final biological endpoint predicted is the cell survival probability, defined as
\begin{equation}\label{EQN:surv}
\mathbb{P}\left ( \lim_{t \to \infty} \left(Y^{(c,1)}(t),\dots,Y^{(c,N_d)}(t)\right) = \mathbf{0} \right)\,,
\end{equation}
\ema{allowing a strict validation as in the following.}

\subsection*{Validation of the MS-GSM$^2$ on \textit{in-vitro} UHDR experimental data}\label{SEC:Val}

\ema{To stress \gsm validity and predictive power,} \fra{the model predictions have been compared with all available experimental data at UHDR for different oxygenation. Such experimental data are of broadly different types, using all the parameters (oxygenation, time structure, doses, radiation quality, etc., as reported in the original publications}
Figure \ref{fig:electron} shows the comparison \fra{between the experimental cell} survival fractions of prostate cancer cell line DU145 \cite{Adrian2020} \fra{(in blue)} and MS-GSM$^2$ \fra{prediction (in black)}. \fra{The horizontal axis} \ema{lists different conditions, not to scale, reporting the different levels of oxygenation (1.6$\%$, 2.7$\%$, 4.4$\%$, 8.3$\%$, 20.0$\%$ $\mathrm{O_2}$) as indicated in the original publication}, for \fra{electron at 10 MeV} in Conv (mean dose rate of 14 Gy/min) and UHDR (mean dose rate of 600 Gy/s) irradiations of 18 Gy. In particular, the dose was delivered with an integer number of pulses, a pulse time of 3.5 $\mu$s, and a pulse repetition frequency of 200 Hz in the case of Conv irradiation, while with an integer number of pulses, a dose-per-pulse of 3 Gy, and a pulse repetition frequency of 200 Hz at UHDR regime. \fra{The experimental irradiation structure has been incorporated within the \gsm framework. As illustrated in Figure \ref{fig:electron}, the \gsm encompasses the OER effect, evidenced by the elevated cell survival fraction at reduced oxygen concentrations for both conventional and FLASH irradiation, as well as the FLASH sparing effect distinguishing the conventional and UHDR regimes.}   

Figure \ref{fig:heliumd8} \ema{reports} the survival fractions of human lung epithelial cells A549 (experimental \cite{Tessonnier2021} results (in blue) \fra{plotted against \gsm prediction (in black)}), at hypoxia and normoxia conditions (1.0$\%$ and 21.0$\%$ $\mathrm{O_2}$), for Conv and UHDR irradiations of 8 Gy, in the case of helium ions \ema{impinging the cells at a LET of}  4.5 keV/$\mu$m. In particular, the dose was delivered with a mean dose rate of 0.12 Gy/s for Conv, with a mean dose rate of 205.13 Gy/s. 

Figure \ref{fig:carbon} shows the comparison of survival fractions of Chinese hamster ovary cells CHO-K1, between \fra{the \gsm prediction (in black)} and the experiment \cite{Tinganelliinvitro2022} (in blue), at different oxygen concentrations (0.5$\%$, 4.0$\%$, 21.0$\%$ $\mathrm{O_2}$). \fra{Carbon ion irradiation has been performed} for both Conv (mean dose rate of 0.6 Gy/s) and UHDR (mean dose rate of 70 Gy/s) irradiations of 7.5 Gy, in the case of carbon ions of 13 keV/$\mu$m. 

Figure \ref{fig:electron_carbon} highlights how the \gsm can describe the trend of the experimental data at different oxygenation levels in terms of oxygenation and sparing at different dose rates and for different radiation quality.

\marco{For each of these experiments, we calibrated the biological parameters of our model, i.e., $a$, $b$, and $r$, on the experimental points at standard conditions, i.e., for conventional irradiation and normoxia, that is 21$\%$ oxygenation. The MS-GSM$^2$ has completely predicted all the other results. We discuss details regarding the calibration procedure in the SI Appendix.}

\subsection*{Translation to \textit{in-vivo}: \textit{in-silico} evidence on a new mechanism for the differential effect}\label{SEC:DifEf}

Figure \ref{fig:XSH_O2} reports the combined action of oxygenation and antioxidant environment on the emergence of the FLASH effect for electron irradiation. In particular, the color map distribution describes the relative variation of the surviving probability for UHDR irradiation with respect to the conventional one, i.e. $(\ln{S_{Conv}}-\ln{S_{FLASH}})/\ln{S_{Conv}}$, as a function of the environmental oxygen (in the x-axis) and the rate of antioxidants (in the y-axis), i.e., the product between the concentration of a chemical species and the reaction constant. We obtained this by repeating the simulation of the irradiation with electrons for several levels of antioxidant concentrations and at different oxygen conditions and considering the same physical and biological parameters that we used for \cite{Adrian2020}. In particular, light colors correspond to a high differential effect between conventional and UHDR irradiation, while dark colors correspond to a low differential effect.

\section*{Discussion}

\fra{Results reported in Section \ref{SEC:Val} show that} MS-GSM$^2$ can describe the empirical trend on a wide range of dose and dose rate values, particle type, and energy, ranging from electrons to carbon ions and environmental oxygen conditions, from severe hypoxia (0.5\% $\mathrm{O_2}$) to normoxia (21\% $\mathrm{O_2}$). \fra{This is achievable due to the remarkable flexibility and comprehensiveness of the \ema{model}, which simultaneously encompasses various physical, chemical, and biological aspects across different spatial and temporal scales.} \marco{In particular, temporally, the MS-GSM$^2$ can span from $10^{-6}$ to $10^{0} \ s$ with the chemical stage and up to $ 10^{5} \ s$ for the DNA damage evolution. Spatially, the MS-GSM$^2$ can describe the physical and chemical events that occur between $10^{-6}-10^{-5} \ m$, up to the macroscopic scale for biology. Everything that happens below these spatial and temporal scales is, however, taken into account by the MS-GSM$^2$ because it is parameterized through a track-structure Monte Carlo approach, for example, to estimate the DNA damage yield in standard conditions, or using an analytical approach, e.g., the amorphous track model for the calculation of the spatial dose distribution. Extensions to include explicit descriptions beyond these temporal and spatial scales are under investigation. However, as demonstrated for electrons, protons, and carbon ions, \cite{baikalov2023intertrack, thompson2023investigating, weber2022flash}, the early stage recombination is supposed to play a minor role.} 

\begin{figure*}[!t]
     \centering
     \includegraphics[width=0.85\textwidth]{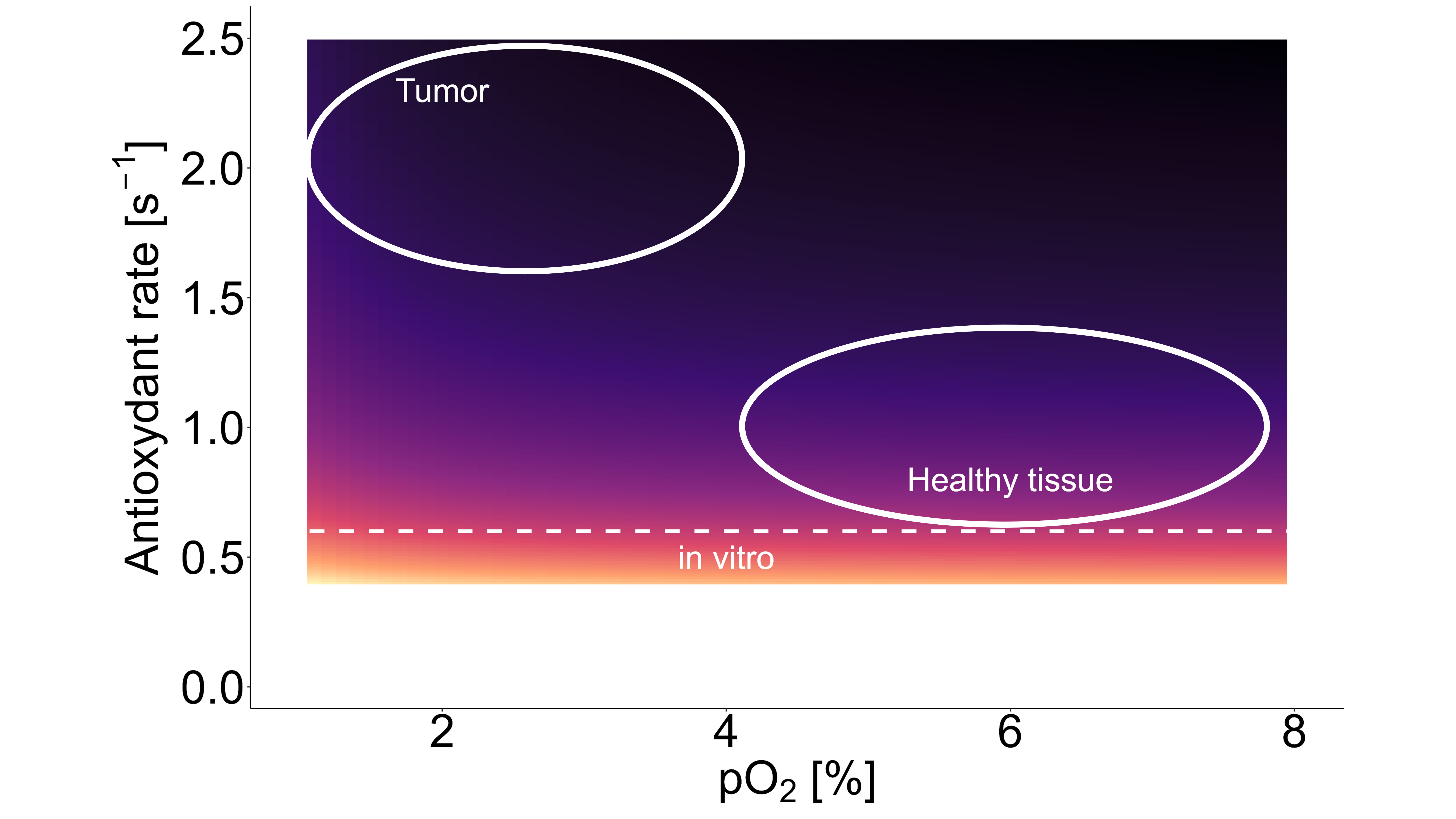}
     \caption{Perspective analysis of the impact of the chemical environment on the emergence of the FLASH effect for electron irradiation. Relative change in cell survival between Conv and UHDR irradiation. Light colors correspond to a high differential effect, while dark colors correspond to a low differential effect. \textit{in-vitro}, \textit{in-vivo} - healthy tissue, and \textit{in-vivo} - tumor regions are highlighted in white.}
     \label{fig:XSH_O2}
\end{figure*}   

\fra{As mentioned in the previous section, the GSM$^2$ parameters have been consistently calibrated to normoxic conditions during conventional irradiation; all other experimental data points are entirely predicted.} We want to underline the remarkable potential of the MS-GSM$^2$, which can predict the sparing effect that occurs at FLASH regime, seen as the rise of the cell survival curves compared to the conventional irradiation, for several levels of oxygenation, \marco{from 0.5\% to 21\% $\mathrm{O_2}$}, different radiation qualities, i.e., electrons, helium, and carbon ions, \marco{and, finally, different dose levels, as shown in Figure \ref{fig:electron_carbon}.} Currently, there are no other mechanistic models capable of simultaneously incorporating all these effects and, consequently, reproducing all the main \textit{in-vitro} experimental data at UHDR regime, i.e., \cite{Adrian2020, Tessonnier2021, Tinganelliinvitro2022}.

\fra{A key advantage of the MS-GSM$^2$ is its ability to account \ema{explicitly} for the actual temporal structure of dose delivery. Optimizing the physical parameters for the clinical application of FLASH irradiation presents significant challenges due to the various parameters affecting the radiation's temporal structure, such as micro- and macro-pulses, repetition frequency, and dose-per-pulse. Existing models, which rely solely on the average approximation of the beam structure, may face a significant limitation due to the current uncertainty regarding which parameters are most influential in the emergence of the FLASH sparing effect.} The \gsm can go beyond these parameterizations, \fra{encompassing in a unique framework} the actual temporal structure of the radiation, i.e., the one experienced by the cell. A clear example of this is the fact that we can describe both the irradiation with electrons, which is characterized by a series of pulses interspersed with pauses \cite{Adrian2020}, and the irradiation with ions, which occurs with a single macro pulse \cite{Tessonnier2021, Tinganelliinvitro2022}.

A very interesting aspect arises from Figure \ref{fig:electron}, namely the fact that the survival level at 1.6 \% $\mathrm{O_2}$ for conventional irradiation is comparable with those at 2.7 \% $\mathrm{O_2}$ for UHDR irradiation. This means that the rise of the surviving fraction in the case of FLASH irradiation, due to the reduction of indirect damage caused by the higher recombination of organic radicals, can \fra{have the same biological effect of a reduction of oxygenation}. \fra{Environmental oxygen appears to play a crucial role in the emergence of the FLASH effect. As equation \eqref{EQN:chemODEs} indicates, oxygen is essential for producing organic radicals. However, it is important to note that this effect is not directly related to oxygen depletion.} In \ema{the initial years of FLASH mechanistic investigations, the prevailing hypothesis was proposing a transient hypoxia induced by oxygen depletion \cite{petersson2020quantitative}. Still, this hypothesis has been widely discredited by both theoretical calculations \cite{Boscolo2021} and experimental evidence \cite{jansen2021does, cao2021quantification}}.

Another crucial point from Figure \ref{fig:electron_carbon} is that the sparing effect, i.e., the \ema{increase} in cell survival at UHDR, is visible for every oxygenation level. \marco{In particular, the FLASH effect is also visible at 21\% $\mathrm{O_2}$ condition, e.g., for electrons irradiation, as in Figure \ref{fig:electron} or, in general, in \cite{Adrian2020, Adrian2021}, but also in the case of carbon ions, as in Figure \ref{fig:carbon}.} In contrast, in the case of \textit{in-vivo} experiments, the FLASH effect occurs only in hypoxia conditions, while it was not observed at physiologic conditions (4-8\% $\mathrm{O_2}$). In addition, unlike what happens \textit{in-vivo}, in the case of \textit{in-vitro} measurements, the sparing effect occurs in both healthy and tumor cells. These two aspects highlight how, most likely, the FLASH effect \textit{in-vivo} is described by the action of at least two combined effects\ema{: a protective effect depending on the dose rate, and another effect modulating its magnitude, depending on the redox environment}. \marco{Indeed, as already mentioned previously, we hypothesize that the sparing effect for UHDR irradiation, which is also visible \textit{in-vitro}, is due to the different recombination of organic radicals, which decreases the toxicity of the cellular environment, and therefore, the action of indirect action of the radiation on the formation of DNA damages, as recently suggested by an experiment on plasmid \cite{Wanstall2024}. At the same time, we speculate that the action of the redox environment, particularly the presence of antioxidants such as thiols, is the basis for the differential effect between healthy and tumor tissue observed for \textit{in-vivo} experiments. This second mechanism, \ema{which may result in leveling out the magnitude of the DNA damage reduction}, is much more complex than the mechanism underlying the sparing effect, which occurs on an intra-cellular level and results in a variation of the indirect DNA damage yield. This latter one may involve a more macroscopic regulation of the redox environment \textit{in-vivo} compared to \textit{in-vitro}.} 

Through the MS-GSM$^2$, we can detail what happens at the cellular level. To describe what happens \textit{in-vivo}, what should happen is that there should be no differences in normoxia between conventional and UHDR irradiation and, at the same time, that there should be no sparing effect in the case of tumor tissues. We hypothesize that the chemical environment, \fra{and in particular the redox environment}, is crucial to explain the FLASH effect at the \textit{in-vivo} level. \fra{It is believed} that there is a difference in the cellular chemical environment between \textit{in-vitro} and \textit{in-vivo}, due to, for example, higher concentrations of antioxidants or, more generally, to the different ability to remove toxic chemical species from the cellular environment. \fra{From Figure \ref{fig:XSH_O2}, it is evident that in the \textit{in-vitro} setting, the cellular environment exists within a gradient region where the differential effect between conventional and UHDR irradiation is observed. However, transitioning from \textit{in-vitro} to \textit{in-vivo}, particularly increasing the antioxidant concentration, there are instances where a sparing effect is not consistently present.} In particular, we can identify the two regions for healthy and cancer tissues, where there is a gradient of the effect, or there is no effect, respectively, also considering that $[\mathrm{O_2}]_{tumor} < [\mathrm{O_2}]_{healthy}$. Indeed, we can distinguish these two areas, e.g., due to the different concentrations of glutathione ($\mathrm{GSH}$) between normal tissue and tumor \cite{Gamcsik2012}. We also hypothesize that the gradient of the sparing effect of the \textit{in-vitro} region explains both the fact that the FLASH effect is observed with different magnitude for the various cell lines and that it emerges for both healthy and cancer cells, e.g., as in \cite{Adrian2021}. From this Figure \ref{fig:XSH_O2}, it is also possible to appreciate how, for a given level of antioxidants, the sparing effect decreases as the oxygen concentration increases, in agreement with the experiment predictions, e.g., \cite{Tinganelliinvitro2022}. Therefore, \fra{we hypothesize that the varying responses of normal and tumor tissues are attributable to distinct redox environments. Given that various tissues and tumors may possess differing concentrations of antioxidants, which are seldom quantified, our hypothesis could account for the inconsistent experimental observations regarding the FLASH effect. Moreover, our analysis suggests a more intricate role than solely oxygenation, which has been the main focus of the community for the past five years. Indeed, our findings indicate that many parameters contribute to the emergence of the sparing effect. For instance, a lower antioxidant level, typically observed in \textit{in-vitro} cells, requires a reduced dose to trigger a sparing effect. Conversely, as antioxidant levels increase, as seen in healthy tissue or even more so in tumors, a significantly higher dose is necessary to induce the effect. This aligns with experimental observations, where \textit{in-vitro} studies demonstrate a pronounced effect at lower doses (around 8 Gy), in contrast to \textit{in-vivo} experiments that require doses exceeding 10 Gy.} 

\marco{We thus provided a novel mathematical framework, able from one side to reproduce the \textit{in-vitro} data at UHDR and to suggest a consistent picture explaining the most relevant features related to the FLASH observations. Among the novel possibilities enabled by this work is applying the MS-GSM$^2$ to investigate specific collections of \textit{in-vivo} data. A major step forward is exploiting the present tool to facilitate the clinical translation of the FLASH radiation therapy. Indeed, the MS-GSM$^2$, in principle, can straightforwardly be used to enable FLASH treatment planning through a specific biologically-driven optimization of the plan, including organ volume effects, such as in \cite{battestini2022including}.}

\section*{Material and Methods}
\subsection*{Model parameters}
Table S1 reports the values of all model parameters, e.g., the reaction rate constants and the initial concentrations of the chemical species considered in our simulation. In contrast,e the fitted biological rates are reported in Table S2 of the SI Appendix.

\subsection*{Computational information and data analyses}
All the simulations, the data fitting, and analyses reported in this work were performed using \textit{Julia}, described in detail in the SI Appendix.

\subsection*{Data, Materials, and Software Availability}
The raw data supporting the conclusions of this article will be made available by the authors without undue reservation. 

\subsection*{Acknowledgment}
This work has been partially supported by the INFN CSN5 project FRIDA, and partially funded by the European Union under NextGenerationEU PRIN 2022 PNRR Prot P2022TX4FE. Walter Tinganelli and Olga Sokol are gratefully acknowledged for providing the raw data from their experiment. Rudi Labarbe, Joao Seco, Sergei Vinogradov, and Amitava Adhikary are thanked for
fruitful discussions.


\cleardoublepage
\bibliographystyle{apalike}
\bibliography{pnas-sample}

\appendix

\section*{Appendix}
\subsection*{Model parameters}

\begin{table}[h]
\centering
\caption{Chemical stage parameters}\label{TAB:chempar}
\begin{tabular}{llr}
\textbf{Parameters} & \textbf{Value/Range} & \textbf{Reference} \\
\midrule
    $k_1$ & $5 \cdot 10^7 ~(M \cdot s)^{-1}$ & \cite{Labarbe2020, Spitz2019} \\
          & & \cite{ Neta1990, Epp1976, Michael1986}\\
    $k_2$ & $10^4 ~(M \cdot s)^{-1}$ & \cite{Labarbe2020, Riahi2010} \\
    $k_3$ & $6.62 \cdot 10^7 ~(M \cdot s)^{-1}$ & \cite{Labarbe2020} \\
    $k_4$ & $10^3 ~(M \cdot s)^{-1}$ & \cite{Labarbe2020} \\
    $k_5$ & $10^9 ~(M \cdot s)^{-1}$ & \cite{Labarbe2020} \\
    $k_6$ & $10^10 ~(M \cdot s)^{-1}$ & \cite{Labarbe2020} \\
    $k_7$ & $4.62 \cdot 10^4 ~(M \cdot s)^{-1}$ & \cite{Labarbe2020} \\
    $k_8$ & $5 \cdot 10^7 ~(M \cdot s)^{-1}$ & \cite{Labarbe2020} \\
    $k_{9}$ & $10^2 ~(M \cdot s)^{-1}$ & \cite{Spitz2019, Wardman2020} \\
    $[\mathrm{O_2}]_0$ & $0-21 ~\%$ & \cite{Adrian2020, Tessonnier2021, Tinganelliinvitro2022} \\
    $[\mathrm{RH}]$ & $1 ~M$ & \cite{Labarbe2020} \\
    $[\mathrm{cat}]$ & $0.08 \cdot 10^{-6} ~M$ & \cite{Labarbe2020} \\
    $[\mathrm{Fe^{2+}}]$ & $8.9 \cdot 10^{-7} ~M$ & \cite{Labarbe2020} \\
    $[\mathrm{XSH}]$ & $6.5 \cdot 10^{-3} ~M$ & \cite{Labarbe2020} \\
\bottomrule
\end{tabular}
\end{table}

\begin{table}[h]
\centering
\caption{Fitted biological rates}\label{TAB:biorates}
\begin{tabular}{llrrr}
\textbf{Experiment} & \textbf{Reference} & $\textbf{a}$ $[1/h]$ & $\textbf{b}$ $[1/h]$ & $\textbf{r}$ $[1/h]$ \\
\midrule
    DU145, e$^-$ & \cite{Adrian2020} & $7.82 \cdot 10^{-3}$ & $1.83 \cdot 10^{-2}$ & $3.23$ \\
    A549, $^4$He & \cite{Tessonnier2021} & $4.70 \cdot 10^{-3}$ & $1.34 \cdot 10^{-2}$ & $4.51$ \\
    CHO-K1, $^{12}$C & \cite{Tinganelliinvitro2022} & $4.21 \cdot 10^{-3}$ & $2.43 \cdot 10^{-2}$ & $3.68$ \\
\bottomrule
\end{tabular}
\end{table}

In Table \ref{TAB:chempar}, we report all the values of the chemical parameters used in our simulations of the system of Ordinary Differential Equations (ODEs) in Eq. (\textcolor{blue}{2}) and the corresponding references.

In Table \ref{TAB:chempar}, we show the biological rates obtained by comparing the results of MS-GSM$^2$ with the experimental data. In particular, the first line corresponds to the results in Figure \textcolor{blue}{2a}, the second line to Figure \textcolor{blue}{2b}, the third line to Figure \textcolor{blue}{2c}.

\subsection*{Computational information and data analyses}

\begin{algorithm}
\caption{MS-GSM$^2$}\label{alg:MSGSM2}
\begin{algorithmic}[1]
    \State Notation:
    \[
    \begin{split}
        &h_{1+4(d-1)}(x,\xi)=rx^d\,,\quad h_{2+4(d-1)}(x,\xi)=ax^d\,,\quad h_{3+4(d-1)}(x,\xi)=bx^d(x^d-1)\,,\\
        & h_{4+4(d-1)}(x,\xi)=\rho(\xi^d) \,,\quad h_{1+4 N_d}(x,\xi)=\dot{d}\,,\quad h(x,\xi) :=  \sum_{i=1}^{1+4 N_d} h_i(x,\xi)\,.
    \end{split}
    \]
    \State start with an initial chemical environment $\xi(t)=\xi_0$, biological system $X(t)=Y(t)=0$ and initial time $t=0$;
    \State set $T_i=0$, for $i=1,\dots,1+4 N_d$;
    \While{$t<T$}
        \ForAll{$h_i$, $i=1,\dots,1+4 N_d$}
            \State generate a random number $R_i \sim U(0,1)$ and set $U_i := \log \frac{1}{R_i}$;
            \State compute $\tau_i$ solving
            \[
            \int_t^{t+\tau_i} h_i(X(s),\xi_s)ds = S_i - T_i\,.
            \]
            \qquad \quad We use the convention that if $h_i(X(t),\xi_s)=0$, then $\tau_i = \infty$;
        \EndFor
        \State select $\tau := \min_{(i=1,\dots,1+4 N_d)} \tau_i$;
        \State select the corresponding rate $h_i$ at which the minimum is attained;
        \State solve the chemical system for $\xi$ in the time interval $[t,t+\tau]$;
        \ForAll{$h_i$, $i=1,\dots,1+4 N_d$}
            \State set 
            \[
            T_i=T_i + \int_t^{t+\tau} h_i(X(s),\xi_s)ds\,,\;
            \]
            \State update $t=t+\tau$;
            \If{$i = 1+4(d-1)$, for some $d=1,\dots,N_d$,}
                \State update $X(t) = X(t) - 1$, $Y(t)=Y(t)$ and $\xi(t) = \xi(t)$;
            \ElsIf{$i = 2+4(d-1)$, for some $d=1,\dots,N_d$,}
                \State update $X(t) = X(t) - 1$, $Y(t)=Y(t)+1$ and $\xi(t) = \xi(t)$;
            \ElsIf{$i = 3+4(d-1)$, for some $d=1,\dots,N_d$,}
                \State update $X(t) = X(t) - 2$, $Y(t)=Y(t)+1$ and $\xi(t) = \xi(t)$;
            \ElsIf{$i = 4+4(d-1)$, for some $d=1,\dots,N_d$,}
                \State simulate $Z_{Y;i}^{(c,d)}$ and $Z_{X;i}^{(c,d)}$; 
                \State update $X(t) = X(t) + Z_{X;i}^{(c,d)}$, $Y(t)=Y(t)+Z_{Y;i}^{(c,d)}$ and $\xi(t) = \xi(t)$;
            \ElsIf{$i = 4 N_d$,}
                \State simulate the position where the track hits;
                \State compute the total specific energy deposition $z$ according to microdosimetric spectra;
                \For{$d=1$ to $N_d$}
                    \State compute the specific energy deposition $z^d$ according to an amorphous track model;
                    \State simulate $Z_{Y;d}^{(c,d)}$ and $Z_{X;d}^{(c,d)}$ according to equation \eqref{EQN:DirDam};
                    \State update $X(t) = X(t) + Z_{X;d}^{(c,d)}$, $Y(t)=Y(t)+Z_{Y;d}^{(c,d)}$ and $\xi(t) = \xi(t)+ G_\xi \rho z^d$;
                \EndFor
            \EndIf 
        \EndFor
    \EndWhile
    \State end the system.
\end{algorithmic}
\end{algorithm}

All the simulations, the data fitting, and analyses reported in the main text were performed using \textit{Julia}. In Algorithm \ref{alg:MSGSM2}, we show the stepwise construction of the MS-GSM$^2$, that is based on a \textit{Stochastic Simulation Algorithm} (SSA) \cite{Wei}. It allows us to solve the system in Eq. (\textcolor{blue}{7}).

We use a \textit{Rosenbrock method}, in particular the \textit{Rodas4} solver of the \textit{DifferentialEquations.jl} package, for the numerical integration of the chemical system of ODEs, described in Eq. (\textcolor{blue}{2}), that is coupled with the parameters reported in Table \ref{TAB:chempar}, and with all initial conditions set to zero, except to the oxygen, as shown in Table \ref{TAB:chempar}. This assumption is because while oxygen is an environmental chemical species, all the others are directly produced by radiation or the reactions between chemical species.

For the comparison of the MS-GSM$^2$ results with the $\textit{in vitro}$ experiments, shown in Figure \textcolor{blue}{2}, we optimize the biological rates, reported in Table \ref{TAB:chempar}, on experimental data at standard conditions, i.e., conventional irradiation, and normoxia (21\% $\mathrm{O_2}$). In particular, we use a cross-entropy method to obtain the best combination of $a$, $b$, and $r$ rates, allowing us to describe all the experimental points. We start from an interval $[l_j, u_j]$ of possible realistic values for these biological rates $j \in {a, b, r}$, i.e. $a \in [10^{-3}, 10^{-1}]~1/h$, $b \in [10^{-3}, 10^{-1}]~1/h$, and $r \in [2.0, 5.0]~1/h$. We randomly sample $10^3$ values of $\log{a}$, $\log{b}$, and $\log{r}$ from uniform distributions $[\log{l_j}, \log{u_j}]$. We compute the normalized mean square error, i.e. $\text{NMSE} = \mathrm{mean} \left( \left( S_{exp} - S_{MS-GSM^2} \right)^2 / S_{exp} \right)$, as proposed in \cite{Missiaggia2024}, and we consider the best $50$ values, that we call $\text{NMSE}_{50}$. We calculate the biological rates $l_{j,min} \left( \min \left( \text{NMSE}_{50} \right) \right)$ and $u_{j,max} = j \left( \max \left( \text{NMSE}_{50} \right) \right)$, with $j \in {a, b, r}$. We consider the new intervals, i.e. $[\log{l_{j,min}}, \log{u_{j,max}}]$, and we repeat N-times the previous procedure, selecting, in the end, the best combination of $a$, $b$, and $r$.

As discussed in the main text, the creation process of indirect DNA damages is described by a Poisson random variable of average $\kappa_I \left( \mathrm{LET}, [\mathrm{O_2}], [\mathrm{ROO^\bullet}]_t \right) = \kappa\left( \mathrm{LET}, [\mathrm{O_2}] \right)\varrho \int_{0}^t [\bar{\mathrm{ROO^\bullet}}](s) \,ds$. The modulation term $\varrho \int_{0}^t [\mathrm{ROO^\bullet}](s) \,ds$ of DNA lesions is normalized to 1, on average over the nucleus, to be consistent with the DNA damage yield at standard conditions \cite{kundrat2020analytical}, i.e., conventional irradiation and 21 \% $\mathrm{O_2}$. At the same time, it naturally decreases as the oxygenation level decreases and the dose rate increases, in such a way that the average damage yield $\kappa \left( \mathrm{LET}, [\mathrm{O_2}], [\mathrm{ROO^\bullet}]_t \right)$ is in agreement with the Oxygen Enhancement Ratio (OER) formulation \cite{Scifoni2013, Epp1972} for conventional irradiation.

\end{document}